\documentclass[11pt,letterpaper]{article}
\pdfoutput=1
\usepackage{jheppub}
\usepackage{color}
\usepackage{graphicx}

\usepackage[english]{babel}

\usepackage{floatrow}

\newfloatcommand{capbtabbox}{table}[][\FBwidth]

\usepackage{blindtext}

\usepackage{verbatim}
\usepackage{amsmath}
\usepackage{amssymb}
\usepackage{subfigure}
\usepackage{url}

\usepackage{multirow}
\usepackage{threeparttable}
\usepackage{paralist}

\newcommand{\GeV}{\text{GeV}}
\newcommand{\TeV}{\text{TeV}}

\newcommand{\U}{\text{U}}
\newcommand{\SU}{\text{SU}}
\newcommand{\SO}{\text{SO}}

\newcommand{\M}{\widetilde{M}}
\newcommand{\m}{\widetilde{m}}
\newcommand{\MG}{\widetilde{M}_{\widetilde{g}}}
\newcommand{\MW}{\widetilde{M}_{\widetilde{W}}}
\newcommand{\MB}{\widetilde{M}_{\widetilde{B}}}
\newcommand{\At}{A_{H_u \widetilde{t}_L \widetilde{t}_R}}
\newcommand{\Stop}{\widetilde{t}}
\newcommand{\Tr}{\texrm{Tr}}

\newcommand{\vev}[1]{\langle #1 \rangle}

\DeclareRobustCommand{\Sec}[1]{Sec.~\ref{#1}}

\DeclareRobustCommand{\App}[1]{App.~\ref{#1}}
\DeclareRobustCommand{\Tab}[1]{Table~\ref{#1}}

\DeclareRobustCommand{\Fig}[1]{Fig.~\ref{#1}}

\DeclareRobustCommand{\Eq}[1]{Eq.~(\ref{#1})}
\DeclareRobustCommand{\Eqs}[2]{Eqs.~(\ref{#1}) and (\ref{#2})}
\DeclareRobustCommand{\Ref}[1]{Ref.~\cite{#1}}
\DeclareRobustCommand{\Refs}[1]{Refs.~\cite{#1}}

\newcommand{\be}{\begin{equation}}
\newcommand{\ee}{\end{equation}}

\newcommand{\mb}[1]{\boldsymbol{#1}}

\newcommand{\kahler}{K\"{a}hler }

\def\Tr{\mathop{\rm Tr}}

\begin{document}

\title{Auxiliary Gauge Mediation: \\ A New Route to Mini-Split Supersymmetry}
 
\author[]{Yonatan Kahn,}
\author[]{Matthew McCullough,}
\author[]{and Jesse Thaler}

\affiliation[]{Center for Theoretical Physics, Massachusetts
  Institute of Technology,\\Cambridge, MA 02139, USA}

\emailAdd{ykahn@mit.edu}
\emailAdd{mccull@mit.edu}
\emailAdd{jthaler@mit.edu}

\date{\today}

\abstract{The discovery of a standard-model-like Higgs at 126 GeV and the absence of squark signals thus far at the LHC both point towards a mini-split spectrum for supersymmetry.  Within standard paradigms, it is non-trivial to realize a mini-split spectrum with heavier sfermions but lighter gauginos while simultaneously generating Higgs sector soft terms of the correct magnitude, suggesting the need for new models of supersymmetry breaking and mediation.  In this paper, we present a new approach to mini-split model building based on gauge mediation by ``auxiliary groups'', which are the anomaly-free continuous symmetries of the standard model in the limit of vanishing Yukawa couplings.  In addition to the well-known flavor $\SU(3)_F$ and baryon-minus-lepton $\U(1)_{B-L}$ groups, we find that an additional $\U(1)_H$ acting on the Higgs doublets alone can be used to generate Higgs soft masses and $B$-terms necessary for a complete model of mini-split.  Auxiliary gauge mediation is a special case of Higgsed gauge mediation, and we review the resulting two-loop scalar soft terms as well as three-loop gaugino masses. Along the way, we present a complete two-loop calculation of $A$-terms and $B$-terms in gauge mediation, which---contrary to a common misconception---includes a non-zero contribution at the messenger threshold which can be sizable in models with light gauginos.  We present several phenomenologically acceptable mini-split spectra arising from auxiliary gauge mediation and highlight a complete minimal model which realizes the required spectrum and Higgs sector soft terms with a single $\U(1)_X$ auxiliary gauge symmetry.  We discuss possible experimental consequences.}


\arxivnumber{}

\preprint{MIT-CTP 4475}

\maketitle

\section{Introduction}
\label{sec:introduction}

The discovery of a 126 GeV Higgs boson \cite{Aad:2012tfa,Chatrchyan:2012ufa} places considerable restrictions on supersymmetry (SUSY) model building. The heavy top squark required for such a large Higgs mass \cite{Martin:1997ns}---combined with flavor bounds and the desire to preserve gauge coupling unification \cite{Dimopoulos:1981yj, Dimopoulos:1981zb}---have pointed to models of mini-split SUSY \cite{Arvanitaki:2012ps,ArkaniHamed:2012gw}, a version of split supersymmetry \cite{Wells:2003tf,Giudice:2004tc,ArkaniHamed:2004fb} where the scalar superpartners are heavier than the gauginos, but not arbitrarily so.\footnote{For other models realizing a similar spectrum, see \Refs{Hall:2012zp,Arganda:2013ve,Baryakhtar:2013wy,Ibanez:2013gf,Grajek:2013ola,McKeen:2013dma,Eliaz:2013aaa,Sato:2013bta}.} Indeed, the Higgs mass in mini-split models forces the third generation squarks to be between 1 and $10^5$ TeV, depending on $\tan \beta$ \cite{ArkaniHamed:2012gw, Arvanitaki:2012ps}. 

With quasi-decoupled squarks evading many experimental bounds, arguably the strong\-est constraint on mini-split models come from the theoretical challenge of obtaining the correct standard model (SM) vacuum structure.\footnote{Of course, there are also constraints if one chooses to require a suitable dark matter candidate with the correct relic density.}  The light gluino does not protect top squarks from running tachyonic under renormalization group (RG) flow, often leading to unacceptable color- and charge-breaking vacua \cite{Ibarra:2005vb,Arvanitaki:2012ps}.  This problem is exacerbated by two-loop RG effects if the first- and second-generation squarks are split from the third \cite{ArkaniHamed:1997ab,Agashe:1998zz}.  Any complete model of mini-split must also generate appropriate Higgs sector soft terms $m^2_{H_u}$, $m^2_{H_d}$, and, most acutely, $B_\mu$.  Of course, mini-split models always include some degree of fine-tuning of parameters to get the correct vacuum, but even to begin fine-tuning, the Higgs soft terms must be at least ``in the ballpark'', which in this context means a value of $\sqrt{B_\mu}$ close to the scalar mass scale.  Thus, mini-split model building is not as simple as ``heavy sfermions, light gauginos'', since one must also ensure the consistency of the Higgs sector. 
 
In this paper, we present a new approach for mini-split model building, which we dub \emph{auxiliary gauge mediation}. In any incarnation of gauge mediation, one is already committed to introducing scales intermediate between the weak scale and the Planck scale (at minimum, the messenger scale), so it is attractive to entertain the possibility of new gauge groups which are spontaneously broken at high scales.  We consider gauging $G_{\rm aux}$, the \emph{auxiliary group} containing all anomaly-free continuous symmetries of the SM in the limit of vanishing Yukawas, consistent with grand unified theories (GUTs).\footnote{By ``anomaly-free'' we mean that $G_{\rm aux}$ has no mixed anomalies with SM gauge groups. $G_{\rm aux}$ may have its own internal anomalies whose cancellation requires the addition of new matter, but these new fields have no SM gauge charges.}  As we will show,
\be
G_{\rm aux} \equiv \SU(3)_F \times \U(1)_{B-L} \times \U(1)_H,
\ee
which contains an $\SU(3)_F$ flavor symmetry that rotates the three generations, the well-known $\U(1)_{B-L}$ symmetry, and most importantly a $\U(1)_H$ symmetry acting on the Higgs doublets.\footnote{A similar $\U(1)_H$ was discussed in \Ref{Ko:2012hd} in the context of non-supersymmetric two-Higgs-doublet models.} Gauge mediation via this spontaneously-broken $\U(1)_H$ generates precisely the Higgs sector soft terms one needs for consistent mini-split model building.  Furthermore, auxiliary gauge mediation ensures that gaugino masses stay two loop factors smaller than scalar masses, automatically realizing the mini-split spectrum.

Auxiliary gauge mediation is a special case of Higgsed gauge mediation \cite{Gorbatov:2008qa}, and we review how to obtain the spectrum at lowest order in the SUSY-breaking parameter $F$ using the techniques of \Refs{Craig:2012yd,Craig:2012di}.  We also present for the the first time a Feynman diagrammatic calculation of the two-loop contribution to $A$- and $B$-terms to all orders in $F$ in Higgsed gauge mediation, which also sheds light on the two-loop result in standard gauge mediation \cite{Rattazzi:1996fb}.  Contrary to a common misconception, we find two-loop contributions to $A$- and $B$-terms which are non-zero at the messenger scale, in addition to the well-known contributions proportional to $\log(M/\overline{\mu})$ which vanish when the RG scale $\overline{\mu}$ equals the messenger scale $M$.\footnote{The bar on $\overline{\mu}$ emphasizes that throughout this paper, we work in the dimensional reduction scheme $\overline{\text{DR}}$.  This is particularly relevant for the discussion in \Sec{sec:standardgauge}, where we want to track finite two-loop contributions. In an earlier calculation \cite{Rattazzi:1996fb}, these contributions were absorbed into a redefinition of the messenger scale.}  Our result is consistent with the known results from analytic continuation into superspace \cite{Giudice:1997ni,ArkaniHamed:1998kj}, where logarithmically-enhanced two-loop $A$- and $B$-terms arise from one-loop RG evolution.  The two-loop contributions we find are not logarithmically-enhanced and therefore a small effect in standard gauge mediation.  They are important, however, to include when studying mini-split models where visible-sector gaugino-loop contributions to $B_\mu$ are suppressed.

For mini-split model building, auxiliary gauge mediation exhibits a number of interesting features.  For concreteness, we will keep our discussion within the context of the minimal supersymmetric standard model (MSSM \cite{Dimopoulos:1981zb}), though auxiliary gauge mediation could be adapted to non-minimal scenarios as well.\footnote{In the context of the next-to-miminal supersymmetric standard model (NMSSM), it would be interesting to augment $G_{\rm aux}$ with additional $\U(1)$ symmetries acting on the singlet superfield.}
\begin{itemize}
\item  While only $\SU(3)_F$ contributes to the gluino soft mass, all three factors in $G_{\rm aux}$ contribute to the wino and bino soft masses.  This allows the gaugino spectrum to be significantly altered relative to more conventional scenarios.  In particular, using the $U(1)_H$ factor, the wino or bino could be close in mass to (or possibly heavier than) the gluino.
\item  The spontaneous breaking of $\SU(3)_F$ allows splittings between the third-generation squarks and those of the first two generations.  This can significantly enhance the branching ratio of gluino decays into third-generation quarks, leading to ``flavored'' mini-split LHC signatures.
\item  Because of the $\U(1)_{B-L}$ factor, auxiliary gauge mediation can accommodate scenarios with sleptons significantly heavier than squarks.
\item As is typical in gauge mediation, the gravitino is the LSP, but generic low-scale models have gravitinos which are too light to be dark matter. Auxiliary mediation using all three factors of $G_{\rm aux}$ can provide a low-scale mini-split spectrum with super-WIMP  \cite{Feng:2003xh, Feng:2003uy} gravitino dark matter, thanks to a bino NLSP of the correct mass.
\item  Economical models of mini-split can be constructed based on the single gauge symmetry $\U(1)_{B-L+k H}$, where $k$ encodes the freedom to choose a variety of Higgs charges. These ``minimal mini-split'' models generate novel, testable gaugino spectra, as well as the necessary Higgs sector soft terms.

\end{itemize}

The structure of this paper is as follows. In \Sec{sec:higgsedmediation}, we review the mechanism of Higgsed gauge mediation for a general gauge group $G$, giving expressions at lowest non-trivial order for all the soft terms.  We take a short detour in \Sec{sec:standardgauge} and \App{app:allorders}, calculating the $A$- and $B$-terms for the case of standard gauge mediation and demonstrating the presence of non-zero contributions at the messenger scale.  \Sec{sec:auxmediation} motivates and defines the auxiliary group $G_{\rm aux}$ and contains the main technical results of our paper.  We provide example spectra and consider associated phenomenology in \Sec{sec:examples}, including scenarios with and without flavor structure. We describe a minimal $\U(1)_{B-L+k H}$ benchmark model in \Sec{sec:minimal}, and conclude in \Sec{sec:conclusions}.

\section{Review of Higgsed Gauge Mediation}
\label{sec:higgsedmediation}

Before studying auxiliary gauge mediation in particular, we first review the broad features of Higgsed gauge mediation.  The reader familiar with this material and the notation in \Ref{Craig:2012yd} can safely skip to \Sec{sec:standardgauge}.

\subsection{Soft Masses from the Effective K\"ahler Potential}
\label{sec:susybreaking}

In Higgsed gauge mediation \cite{Gorbatov:2008qa}, SM soft masses arise from messengers charged under a spontaneously broken gauge symmetry.  For simplicity, consider an Abelian gauge group $\U(1)'$ and a single vector-like messenger $\mb{\Phi},\mb{\Phi^c}$ with charge $q_\Phi$.  As in minimal gauge-mediated scenarios, these messengers are coupled to the SUSY-breaking spurion $\langle \mb{X} \rangle = M + \theta^2 F$ in the superpotential
\be
W \supset \mb{X \Phi \Phi^c}.
\ee
The generalization to non-Abelian gauge groups and multiple messengers is straightforward.

Because $\U(1)'$ is spontaneously broken at a high scale, the calculation of soft-masses is considerably more complicated than for standard gauge mediation, and the elegant technique of analytically-continuing RG thresholds \cite{Giudice:1997ni,ArkaniHamed:1998kj} cannot be directly employed due to the multiple mass thresholds.  As shown in \Ref{Craig:2012yd} and later applied in \Ref{Craig:2012di}, the full soft spectrum can be obtained by employing the two-loop effective \kahler potential and analytically continuing both the messenger mass and the vector superfield mass,
\be
\label{eq:neededcontinuation}
|\mb{M_\Phi}|^2 \rightarrow \mb{X}^\dagger \mb{X}, ~~~
\mb{M_V}^2 \rightarrow M_V^2+2 {g'}^2 q_q^2 \mb{q}^\dagger \mb{q},
\ee
where $q$ are visible-sector fields with charge $q_q$ under the $\U(1)'$.

Using the two-loop effective \kahler potential result from \Ref{Nibbelink:2005wc} and the two-loop sunrise-diagram integral evaluated in \Ref{Ford:1991hw}, we have
\begin{align}
K_{2L}  & \supset   \frac{q_\Phi^2 g^2}{(4 \pi)^4} |\mb{M_\Phi}|^2 \bigg[ 2 \mb{\Delta}  \log (\mb{\Delta}) \left( \log \left({\tfrac{|\mb{M_\Phi}|^2}{\overline{\mu}^2}}\right) -2 \right)   \nonumber \\
& \quad ~ + (\mb{\Delta}+2) \log \left({\tfrac{|\mb{M_\Phi}|^2}{\overline{\mu}^2}}\right) \left( \log \left({\tfrac{|\mb{M_\Phi}|^2}{\overline{\mu}^2}}\right) -4  \right) + \Omega (\mb{\Delta})  \bigg] , \qquad \mb{\Delta}  \equiv   \frac{\mb{M_V}^2}{|\mb{M_\Phi}|^2} ,
\label{eq:int2}
\end{align}
where $\overline{\mu}$ is the $\overline{\text{DR}}$ renormalization scale, and we can express the function $\Omega(\mb{\Delta})$ using dilogarithms as
\be
\Omega (\mb{\Delta}) = \sqrt{\mb{\Delta} (\mb{\Delta}-4)} \left(2 \zeta (2) + \log^2 \left(\mb{\alpha} \right) + 4 \text{Li}_2 \left[-\mb{\alpha} \right] \right) \quad\text{with} \quad \mb{\alpha} = \left(\sqrt{\frac{\mb{\Delta}}{4}} +\sqrt{\frac{\mb{\Delta}}{4} -1} \right)^{-2}.
\label{eq:omega}
\ee
Applying the shift in \Eq{eq:neededcontinuation} and expanding \Eq{eq:int2} to first order in $|\mb{q}|^2$ and lowest non-trivial order in $F/M^2$, we are left with a two-loop \kahler potential for the visible-sector fields
\be
K_{2L}   \supset - q_\Phi^2 q_q^2 \frac{\alpha^2}{(2 \pi)^2} \left(h (\delta) \left(\frac{F}{M} \theta^2 + \frac{F^\dagger}{M^\dagger} \overline{\theta}^2 \right)+ f (\delta) \left| \frac{F}{M} \right|^2 \theta^2  \overline{\theta}^2 \right) |\mb{q}|^2,  \qquad \delta = \left| \frac{M_V}{M} \right|^2,
\label{eq:susykahl}
\ee
where the factors $h(\delta)$ and $f(\delta)$ track the difference between Higgsed gauge mediation and standard gauge mediation,\footnote{For a generalization of the function $h(\delta)$ to all orders in $F/M^2$ see \App{app:allorders}, and for a similar generalization of $f(\delta)$ see \Ref{Gorbatov:2008qa}.} and are given explicitly by
\begin{align}
h(\delta) &= 2 \frac{(\delta-4) \delta \log(\delta)-\Omega(\delta)}{\delta (\delta-4)^2},
\label{eq:hdelta} \\
f(\delta) &= 2 \frac{\delta (\delta-4) ((\delta-4) +(\delta+2) \log(\delta))-2 (\delta-1) \Omega(\delta)}{\delta (\delta-4)^3}.
\label{eq:fd}
\end{align}
From \Eq{eq:susykahl}, we will derive two-loop scalar mass-squared, two-loop $A$- and $B$-terms, and three-loop gaugino masses in the subsections below.

As expected, the SUSY breaking contributions vanish as $\delta \to \infty$ since the gauge superfield becomes infinitely massive and no longer mediates SUSY breaking.  This can be seen from the limiting behavior
\be
\lim \limits_{\delta \to \infty} h(\delta) = \frac{2 \log \delta}{\delta}, \qquad \lim \limits_{\delta \to \infty} f(\delta) = \frac{2(\log \delta -1)}{\delta}.
\ee
The unbroken limit $\delta \to 0$ corresponds to standard gauge mediation,
\be
\label{eq:hlimitdeltazero}
\lim \limits_{\delta \to 0} h(\delta) = (1 - \log \delta), \qquad \lim \limits_{\delta \to 0} f(\delta) = 1.
\ee
Note the large logarithm in $h(\delta)$, corresponding to the $\theta^2$ components in \Eq{eq:susykahl}, which arises from the running of the gauge coupling between the messenger scale $M$ and the vector mass scale $M_V$.   We will return to this function in some detail in \Sec{sec:standardgauge}.

\subsection{Two-Loop Scalar Masses}
\label{sec:scalarmasses}

When the mediating gauge group is Abelian, we can read off the scalar soft mass-squared directly from \Eq{eq:susykahl}:
\be
\widetilde{m}_q^2 = q_q^2 q_\Phi^2 \frac{\alpha^2}{(2 \pi)^2}  \left | \frac{F}{M} \right |^2  f(\delta) , \qquad \delta \equiv \left( \frac{{M_{V}}}{M} \right)^2,
\label{eq:absoftgen}
\ee
where $M_{V}$ is the mass of the $\U(1)'$ gauge superfield, $\alpha = g^2/4 \pi$ is the corresponding fine-structure constant, and $\mb{q}$ and $\mb{\Phi}$ have respective charges $q_q$ and $q_\Phi$.  It is straightforward to generalize to the non-Abelian case \cite{Craig:2012yd},
\be
\left(\widetilde{m}_q^2 \right)_{ij} = C(\mb{\Phi}) \frac{\alpha^2}{(2 \pi)^2}  \left | \frac{F}{M} \right |^2  \sum_a f(\delta^a) \, (T_q^a T_q^a)_{\{ij\}}, \qquad \delta^a \equiv \frac{{M^a_V}^2}{M^2},
\label{eq:nonabgen}
\ee
where $M^a_V$ is the mass of the gauge superfield corresponding to the generator $T^a$, $\{ij\}$ indicates that these indices have been symmetrized and $C(\mb{\Phi})$ is the Dynkin index of the messenger superfield representation. Generalizing to multiple gauge groups and multiple messengers is more complicated if the gauge groups mix (see \Ref{Craig:2012yd}).  We will consider scenarios where mixing is not present in this paper for simplicity of presentation, in which case we need only include a sum over various messenger/gauge group contributions. 

The formul\ae \ in \Eqs{eq:absoftgen}{eq:nonabgen} are the values of the soft masses at the \emph{effective} messenger scale, which is the lower of the scales $M$ or $M_V$.  Specifically, if the gauge symmetry is spontaneously broken far below the messenger scale $M$, the effective messenger scale is $M_V$ rather than $M$ since the ``running'' from the scale $M$ down to $M_V$ has already been accommodated by the effective \kahler potential.\footnote{Strictly speaking, the effective \kahler potential does not include resummation of logarithms, but this prescription for the effective messenger scale is needed to avoid double-counting of the momentum scales between $M$ and $M_V$.}  Hence, the proper definition of the effective messenger scale $M_{\rm eff} = \min\{M,M_V\}$ is important when RG-evolving the soft terms from high scales down to the weak scale through their interactions with the visible sector.

\subsection{Two-Loop A-terms and B-terms}
\label{sec:bmu}

To find the two-loop $A$- and $B$-terms, it is easiest to holomorphically rescale each visible-sector superfield to eliminate terms linear in $\theta^2$ in \Eq{eq:susykahl}:
\be
\mb{q} \rightarrow \left(1+ q_q^2 q_\Phi^2 \frac{\alpha^2}{(2 \pi)^2} h (\delta) \frac{F}{M} \theta^2 \right) \mb{q} ,
\label{eq:rescaleab}
\ee
or in the non-Abelian case
\be
\mb{q_i} \rightarrow \left(\delta_{ij}+ C(\mb{\Phi}) \frac{\alpha^2}{(2 \pi)^2} \sum_a h(\delta^a) \, (T_q^a T_q^a)_{\{ij\}} \frac{F}{M} \theta^2 \right) \mb{q_j} .
\label{eq:rescalenonab}
\ee
This rescaling does not affect the value of the soft masses at two-loop order since the resulting corrections appear formally at four loops.  With this holomorphic rescaling, the SUSY breaking terms are pulled into the superpotential, leading to SUSY-breaking holomorphic terms in the scalar potential.

Adapting the notation of \Ref{Giudice:1997ni}, we can write the soft scalar potential as
\be
\label{eq:definingAi}
V_{\rm soft} \supset \sum_{ij} A_{ij} \tilde{q}_i \frac{\partial \mb{W}}{\partial \mb{q_j}} \bigg|_{\theta^2 \rightarrow 0}.
\ee
In the Abelian case we have 
\be
\label{eq:Vholab}
A_{ij} = A_i \delta_{ij}, \qquad A_i = q_q^2 q_\Phi^2 \frac{\alpha^2}{(2 \pi)^2} h (\delta) \frac{F}{M},
\ee
and in the non-Abelian case
\be
\label{eq:Vholnonab}
A_{ij} = C(\mb{\Phi}) \frac{\alpha^2}{(2 \pi)^2} \frac{F}{M}  \sum_a h(\delta^a) (T_q^a T_q^a)_{\{ij\}}.
\ee
Again, these soft terms should be considered to appear at the effective messenger scale $M_{\rm eff} = \min \{M_V, M \}$.  In \Sec{sec:standardgauge}, we will discuss how to interpret the $M_V \to 0$ limit.

\subsection{Three-Loop Gaugino Masses}
\label{sec:gauginomasses}

If the messengers $\mb{\Phi},\mb{\Phi^c}$ are uncharged under SM gauge groups, then visible-sector gaugino masses first arise at three-loop order.  Though this might seem computationally daunting, one can again use the power of holomorphy and analytic continuation to extract this three-loop effect from \Eq{eq:susykahl}.  The field rescaling in \Eqs{eq:rescaleab}{eq:rescalenonab} is anomalous \cite{Clark:1979te,Konishi:1983hf}, leading to a shift of the gauge kinetic function
\be
\int d^2 \theta \, \mb{f} \, \mb{W}_\alpha \mb{W}^\alpha \to \int d^2 \theta \left(\mb{f}  - \sum_{\mb{q}_r} \frac{C_G (\mb{q}_r)}{8 \pi^2} \log \mb{Z}_{\mb{q}_r} (\overline{\mu})  \right) \mb{W}_\alpha \mb{W}^\alpha.
\ee
Since this rescaling contains SUSY-breaking components, it leads to Majorana gaugino masses.\footnote{For a discussion of how this effect can be seen from the point of view of the real gauge coupling superfield, see \Refs{Giudice:1997ni,ArkaniHamed:1998kj,Craig:2012di}.}

If the visible-sector chiral superfields $\mb{q}_r$ are charged under an Abelian mediating gauge group, then the gaugino mass for a visible-sector gauge group $G$ is
\be
\label{eq:gauginomassab}
\widetilde{M}_{\lambda_G} = q_\Phi^2 \frac{\alpha_G}{2 \pi}  \frac{\alpha^2}{(2 \pi)^2} h (\delta) \frac{F}{M} \sum_{\mb{q}_r} q_q^2 C_G (\mb{q}_r),
\ee
where the sum is over all rescaled fields.  For a non-Abelian mediating gauge group $G'$,
\be
\label{eq:gauginomassnonab}
\widetilde{M}_{\lambda_G} = C(\mb{\Phi})  \frac{\alpha_G}{2 \pi} \frac{\alpha^2}{(2 \pi)^2} \frac{F}{M} \sum_{\mb{q}_r} C_G (\mb{q}_r) C_{G'}  (\mb{q}_r) \sum_{a} h(\delta^a).
\ee
Here the sum over the generators appearing in \Eq{eq:rescalenonab} simplifies using $\Tr(T^a T^b) = C_{G'}\delta^{ab}$, hence the appearance of the Dynkin index of $\mb{q}_r$ with respect to the mediating group $G'$. This simplification still holds even after an orthogonal rotation of the generators $T^a$ to the mass eigenstate basis, since the Dykin index is just the magnitude of $T^a$ with respect to the trace norm.

\section{A-terms and B-terms in Standard Gauge Mediation}
\label{sec:standardgauge}

Before applying the above expressions to the case of auxiliary gauge mediation, it is worthwhile to pause and consider the $\delta \to 0$ limit in more detail, since this should yield the familiar results of standard gauge mediation where the mediating gauge group $G\equiv G_{\rm SM}$ is unbroken.\footnote{Of course, the three-loop gaugino masses in \Sec{sec:scalarmasses} are subdominant in the standard gauge mediation case where gaugino masses first arise at one-loop order, whereas the three-loop gaugino mass is the desired leading effect in auxiliary gauge mediation to get light gauginos in mini-split SUSY.}  Because $f(\delta \to 0) = 1$, the two-loop scalar soft-masses in \Sec{sec:scalarmasses} clearly match those for standard gauge mediation.  At first glance, the $A$- and $B$-term results in \Sec{sec:bmu} also appear to match the standard gauge-mediated results if we reinterpret the vector mass $M_V$ as the RG scale $\overline{\mu}$ and take $h(\delta) \simeq - \log \delta \simeq \log (M^2/\overline{\mu}^2)$.  Indeed, this logarithmic factor is a well-known one-loop effect of RG evolution driven by the gaugino masses.

Upon closer inspection, however, there appears to be a mismatch between the standard lore about $A$- and $B$-terms in gauge mediation and our expressions.  Applying the general results found in \Sec{sec:higgsedmediation} to standard gauge mediation, the SM gauge groups are unbroken above the weak scale so the low energy cutoff in the path integral is the SM gaugino mass rather than the gauge superfield mass.  Thus, in the $\delta \to 0$ limit in \Eq{eq:hlimitdeltazero}, we should really make the replacement
\be
\label{eq:correcthdelta}
h(\delta) \to 1 +  \log \left(\frac{M^2}{\overline{\mu}^2} \right),
\ee
where $M$ is the messenger mass and $\overline{\mu}$ is the RG scale which should be ultimately set to the gaugino mass (which by design is close to the weak scale).  From the results in \Sec{sec:bmu}, we therefore find $A, B_\mu \propto (1 +\log (M^2/\overline{\mu}^2))$.  Naively, this seems to be at odds with previous results based on analytic continuation with one-loop threshold RG matching, where $A, B_\mu \propto \log (M^2/\overline{\mu}^2)$ vanishes at the messenger scale \cite{Giudice:1997ni,ArkaniHamed:1998kj}.  In a common misconception, it is often assumed that $A$- and $B$-terms always vanish at the messenger scale in gauge mediation, although this statement is  in fact  only true at one-loop.\footnote{We are not sure where this misconception comes from, since \Refs{Giudice:1997ni,ArkaniHamed:1998kj} only make this statement for the matched one-loop calculation and not as a claim for the full two-loop result, and a two-loop finite contribution had been calculated previously with Feynman diagrams in \Ref{Rattazzi:1996fb}.}

There are two different ways to see why this standard lore is not quite correct.  First, we can revisit the arguments in \Ref{Giudice:1997ni} on analytic continuation to show why threshold matching and one-loop RG running does not yield the complete answer at two-loop order.  The wavefunction renormalization of a visible-sector superfield $\mb{Q}$ is in general a function of the ultraviolet (UV) gauge coupling $\alpha_{\rm UV}$ defined at the cutoff scale $\Lambda$, and the logarithms $L_X = \log (\overline{\mu}^2/|\mb{X}|^2)$ and $L_{\rm UV} = \log (\overline{\mu}^2/\Lambda^2)$, which can be written generally as
\be
\log (Z_Q) = \sum_\ell \alpha_{\rm UV}^{\ell-1} P_\ell (\alpha_{\rm UV} L_X, \alpha_{\rm UV} L_{\rm UV}),
\ee
where $\ell$ is the loop order.  The soft-masses are calculated from
\begin{eqnarray}
\widetilde{m}^2_Q & = & - \frac{\partial^2 \log (Z_Q)}{\partial \log(\mb{X}) \partial \log(\mb{X^\dagger})} \left| \frac{F}{M} \right|^2 \\
& \propto & \sum_\ell \alpha^{\ell+1} (\overline{\mu}) P''_\ell \left(\alpha (\overline{\mu}) L_X \right),
\end{eqnarray}
where in the second line the loop function $P_\ell$ has been differentiated twice.  Thus the $\alpha^2 (\overline{\mu})$ soft-masses can be evaluated simply with the one-loop running $P_1$, which is the beauty of the argument presented in \Ref{Giudice:1997ni}.  However, if we consider the value of $A_Q$ (see \Eq{eq:definingAi}) that enters into $A$- and $B$-terms, we have
\begin{eqnarray}
A_Q & = & - \frac{\partial \log (Z_Q)}{\partial \log(\mb{X})} \frac{F}{M} \\
& \propto & \sum_\ell \alpha^{\ell} (\overline{\mu}) P'_\ell \left(\alpha (\overline{\mu}) L_X \right),
\end{eqnarray}
where now the loop function has only been differentiated once.  Thus, the full $\alpha^2 (\overline{\mu})$ $A$- and $B$-terms require the full two-loop result; one-loop running and matching cannot capture all of the contributions.  Thus, the general arguments of \Ref{Giudice:1997ni} already accommodate a discrepancy between the full two-loop result obtained here and the result obtained from one-loop RG threshold matching.

Second, we can perform a brute force calculation in component fields to show that \Eq{eq:correcthdelta} is the proper replacement.  In \App{app:allorders}, we perform a full two-loop calculation of $A$- and $B$-terms to all orders in $F/M^2$.  For a broken mediating gauge group in \App{app:allordersbroken}, this yields an effective $\tilde{h} (F/M^2,\delta)$, with the expansion
\be
\tilde{h} (F/M^2,\delta) = h(\delta) + \mathcal{O}\left(\frac{F}{M^2}\right),
\ee
in agreement with the answer obtained using our analytic continuation method.  For an unbroken mediating gauge group in \App{app:allordersunbroken}, the two-loop diagram contains an IR divergence.  In this case, if we regulate this divergence with dimensional reduction ($\overline{\text{DR}}$) (following e.g\ Eq.\ (2.21) of \Ref{Martin:2003qz}), we find that $A, B_\mu \propto (1 + \log(M^2/\overline{\mu}^2))$, which is precisely the form arising from the analytic continuation method used here.\footnote{\Ref{Rattazzi:1996fb} also finds a finite piece, though it is a factor of two larger than what we find here.  See \App{app:allordersunbroken} for a more detailed discussion.} This justifies the replacement of $M_V\rightarrow \overline{\mu}$ in the case of an unbroken gauge group, and demonstrates that $M_V$ can be identified with with the $\overline{\text{DR}}$ RG scale $\overline{\mu}$, making a direct connection (and highlighting the discrepancy) with results based solely on threshold matching.\footnote{This result also has implications for the three-loop gaugino mass contributions, since they arise from precisely the same $\theta^2$ terms in the scalar wavefunction renormalization that generate the $A$- and $B$-terms.}

Practically speaking, the difference between the full two-loop answer $A, B_\mu \propto (1 + \log(M^2/\overline{\mu}^2))$ and the lore $A, B_\mu \propto \log (M^2/\overline{\mu}^2)$ has been relatively unimportant up until now since the logarithmic term typically dominates.\footnote{Getting the precise value of $A$ terms is important when appealing to naturalness considerations, though, since non-zero $A$-terms at the messenger scale help push down the stop masses required for a Higgs at 126 GeV by increasing stop mixing.}    In mini-split models, though, the finite piece is more relevant, since visible-sector gaugino masses can be very small and the precise values of Higgs sector parameters such as $B_\mu$ are important.  

\section{Auxiliary Gauge Mediation}
\label{sec:auxmediation}

\begin{figure}[t]
\begin{center}
\includegraphics[height=0.23\textwidth]{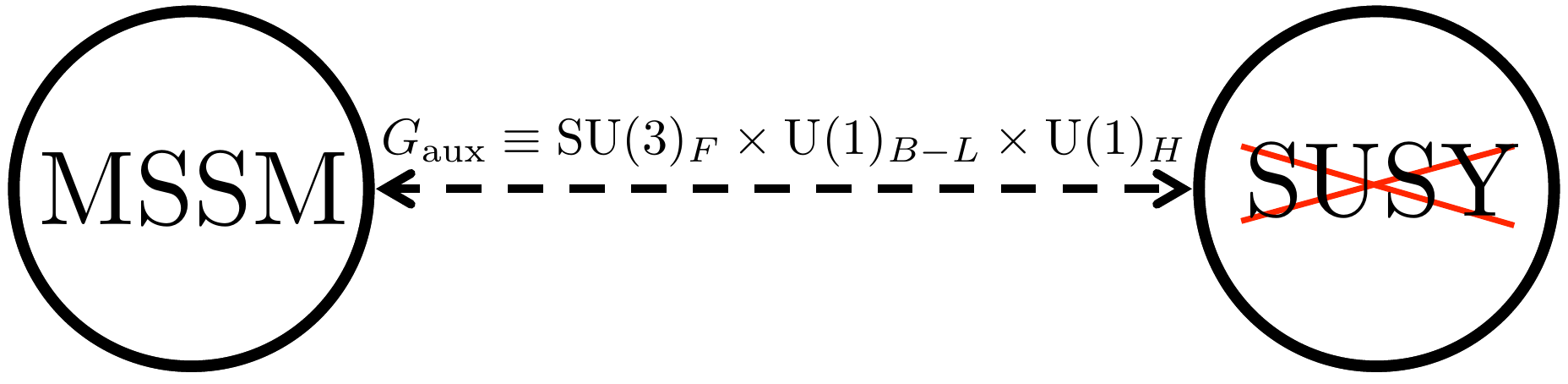}
\caption{General structure of auxiliary gauge mediation, where hidden sector SUSY breaking is communicated to the MSSM via messengers charged only under $G_{\rm aux} \equiv \SU(3)_F \times \U(1)_{B-L} \times \U(1)_H$ (and not under $G_{\rm SM} = \SU(3)_C \times \SU(2)_L \times \U(1)_Y$).}
\label{fig:dumbbell}
\end{center}
\end{figure}

In the framework of auxiliary gauge mediation, SM Yukawa couplings are generated via spontaneous breaking of the auxiliary group 
\be
\label{eq:Gaux}
G_{\rm aux} \equiv \SU(3)_F \times \U(1)_{B-L} \times \U(1)_H
\ee
at a high scale, which we shall refer to as the ``auxiliary scale''.  Above the auxiliary scale, it is consistent for the full gauge group of the MSSM to be
\be
G_{\rm total} \equiv G_{\rm SM} \times G_{\rm aux}, \qquad G_{\rm SM} = \SU(3)_C \times \SU(2)_L \times \U(1)_Y.
\ee
This auxiliary gauge symmetry $G_{\rm aux}$ can then play a role in mediating SUSY breaking to the MSSM fields as shown in \Fig{fig:dumbbell}, leading to new connections between MSSM soft terms and flavor structures.  Gauge mediation by the $\SU(3)_F$ flavor group was previously considered in \Refs{Craig:2012yd,Craig:2012di}, where its role was to augment the contribution from standard $G_{\rm SM}$ gauge mediation.   Here, we will take auxiliary $G_{\rm aux}$ gauge mediation as the sole mediation mechanism, leading to a novel and economical realization of the mini-split SUSY scenario with a (predictive) hierarchy between sfermions and gauginos.

\subsection{Motivating the Auxiliary Group}

Before calculating the soft spectrum, we want to justify the choice of $G_{\rm aux}$ in \Eq{eq:Gaux}.  This can be achieved by switching off the SM Yukawa couplings and considering all possible gauge symmetries consistent with anomaly cancellation.  A powerful simplifying criteria is to require that $G_{\rm aux}$ has no mixed SM gauge anomalies, such that no new SM charged matter is need to cancel anomalies.  This has the appealing feature of not spoiling gauge-coupling unification, though one could of course consider more general gauge groups with exotic matter.

With this criteria imposed, we are left with a small set of possibilities.  In the flavor sector one could have an $\SU(3)_F$ gauge symmetry (with all quark and lepton multiplets transforming in the fundamental) or an $\SO(3)_L \times \SO(3)_R$ gauge symmetry (with the electroweak doublets $\mb{Q}$ and $\mb{L}$ transforming separately from the electroweak singlets $\mb{U^c}$, $\mb{D^c}$, and $\mb{E^c}$).  An $\SO(3)_L \times \SO(3)_R$ gauge symmetry is likely inconsistent with the simplest GUT models, since left-handed and right-handed fields often live in the same GUT multiplets.   For this reason we opt for the $\SU(3)_F$ gauge symmetry in defining $G_{\rm aux}$.\footnote{One could also choose to gauge just an $\SU(2)$ or $\U(1)$ subgroup of the flavor $\SU(3)_F$, acting e.g.\ on the first two generations.   Given that a larger gauge symmetry is possible and there is no obvious reason why only some subgroup would be gauged, we will always gauge the full $\SU(3)_F$.}

Gauge mediation by additional $\U(1)$ gauge groups has been considered before \cite{Kaplan:1998jk,Cheng:1998nb,Cheng:1998hc,Everett:2000hb,Langacker:2007ac,Langacker:2008ip}; all of these models require extra matter with SM gauge charges for anomaly cancellation. An obvious anomaly-free gauge symmetry is $\U(1)_{B-L}$, which has has received considerable attention \cite{Affleck:1984xz,Dobrescu:1997qc,Kikuchi:2008xu}. This, and the $\SU(3)_F$ flavor symmetry, can both be used to generate scalar soft-masses for all of the matter fields.  However gauge mediation by $\SU(3)_F \times \U(1)_{B-L}$ alone leads to issues in the Higgs sector since the Higgs multiplets are uncharged under both mediating groups and, at two loops, Higgs soft-masses squared and the $B_\mu$ term are both vanishing at the messenger scale.  This can be remedied by mixing $\U(1)_{B-L}$ with $\U(1)_Y$ \cite{Mohapatra:1997kv,Arvanitaki:2012ps}, though this option is not in the spirit of this paper where we wish to separate $G_{\rm SM}$ from $G_{\rm aux}$.\footnote{For this case of mixing $\U(1)_{B-L}$ with $\U(1)_Y$, avoiding issues such as tachyonic stops requires the tuning of tree-level $D$-term contributions against two-loop soft masses as well as very particular values of the mixing angle.}

The crucial ingredient for auxiliary gauge mediation is a  $\U(1)_H$ gauge symmetry, under which $H_u$ and $H_d$ have equal and opposite charges and all other fields are neutral.\footnote{Additional anomaly-free $\U(1)$ symmetries acting on Higgs doublets are discussed in \Ref{Ko:2012hd}, but these only apply to the Type I two-Higgs-doublet models, not Type II relevant for SUSY.} This possibility was missed in the first treatment of flavor mediation \cite{Craig:2012di}, though in that context it was relatively unimportant since standard $G_{\rm SM}$ gauge mediation was employed to realize a natural SUSY spectrum.  Here, $\U(1)_H$ is  crucial for successful electroweak symmetry breaking since $\U(1)_H$ leads to Higgs soft-masses and also a $B_\mu$ term at two loops.

\begin{table}[t]
\centering
\begin{tabular}{c || c | c | c || c | c | c} \hline
\hline  & $\SU(3)_C$ & $\SU(2)_{L}$ & $\U(1)_Y$ & $\SU(3)_F$ & $\U(1)_{B-L}$ & $\U(1)_H$\\
\hline \hline $\mb{Q}$ & $\mb{3}$ & $\mb{2}$& $1/6$ & $\mb{3}$ & $1/3$ & ---\\
 $\mb{U^c}$ &$\overline{\mb{3}}$ & --- & $-2/3$ & $\mb{3}$ & $-1/3$ & ---\\
 $\mb{D^c}$ &$\overline{\mb{3}}$ & --- & $1/3$ & $\mb{3}$ & $-1/3$ & ---\\
 $\mb{L}$ & --- & $\mb{2}$ & $-1/2$ & $\mb{3}$ & $-1$ & ---\\
 $\mb{E^c}$ &--- & --- & 1& $\mb{3}$ & $1$ & ---\\
 $\mb{H_u}$ & --- & $\mb{2}$ & $1/2$ & --- & --- & $1$\\
 $\mb{H_d}$ & ---& $\mb{2}$ & $-1/2$ & --- & --- & $-1$\\
\hline $\mb{N_F^c}$ &---&---&---& $\overline{\mb{3}}$ &--- & ---\\
 $\mb{N_{B-L}^c}$ &---&---&---& --- & $1$ & ---\\
 $\mb{S_u}$ &---&---&---& $\overline{\mb{6}}$ & --- &---\\
 $\mb{S_d}$ &---&---&---& $\overline{\mb{6}}$ & --- & ---\\
 $\mb{S_{B-L}^\pm}$ &---&---&---& --- & $\pm 2$ & --- \\
 $\mb{S_H^\pm}$ &---&---&---& --- &--- & $\pm 1$\\
 \hline
 $\mb{\Phi}/\mb{\Phi^c}$ &---&---&---& $ C(\mb{\Phi}) $ & $\pm p_\Phi$ & $\pm q_\Phi$ \\
\hline $\alpha_i$ & $\alpha_S$& $\alpha_W$& $\alpha_Y$& $\alpha_F$ & $\alpha_{B-L}$ & $\alpha_H$\\
\hline\hline
\end{tabular}
\caption{Representations under $G_{\rm total} \equiv G_{\rm SM} \times G_{\rm aux}$ of the MSSM superfields and additional superfields required for anomaly cancellation and the generation of Yukawa couplings.    The notation $C(\mb{\Phi})$ means that the messenger $\mb{\Phi}$ lives in a representation with Dynkin index $C(\mb{\Phi})$. Also shown are the coupling constants $\alpha_i = g_i^2/4\pi$ for the various groups.}
\label{tab:reps}
\end{table}

Thus, we arrive at the most general auxiliary group consistent with the requirements of anomaly cancellation and gauge coupling unification: $G_{\rm aux} \equiv \SU(3)_F \times \U(1)_{B-L}  \times \U(1)_H$. In fact, we may obtain acceptable phenomenology by mediating with $\U(1)_H$ and just one of the other two factors, but in the interest of completeness we will retain this full gauge symmetry in the soft-mass expressions in \Sec{sec:auxsoft}. The representations of the MSSM fields under these gauge symmetries are detailed in \Tab{tab:reps}.  While we have ensured the absence of mixed SM-auxiliary anomalies, additional fields with no SM gauge charges are of course needed to cancel anomalies within $G_{\rm aux}$ itself.  An example of a fully anomaly-free spectrum is given in  \Tab{tab:reps}, motivated by the states needed below to break $G_{\rm aux}$ and generate Yukawa couplings.

\subsection{Flavor Boson Mass Spectrum}
\label{sec:breaking}

In order to calculate soft terms, we need to know some details about the breaking of $G_{\rm aux}$ at the auxiliary scale.  While a complete model of Yukawa coupling generation is beyond the scope of this work, we do need to choose a specific field content and  vacuum expectation value (vev) structure to know the auxiliary gauge boson mass spectrum.  Following \Ref{Craig:2012di} and summarized in \Tab{tab:reps}, we assume that the only sources of $\SU(3)_F$ breaking are fields $\mb{S_u}$ and $\mb{S_d}$ (both transforming as a $\mb{\overline{6}}$ under $\SU(3)_F$), which get vevs along a $D$-flat direction as to not break SUSY. The fields $\mb{S_{B-L}^\pm}$ ($\mb{S_H^\pm}$) are responsible for breaking $\U(1)_{B-L}$ ($\U(1)_H$). The additional right-handed neutrino fields $\mb{N_F^c}$ and $\mb{N_{B-L}^c}$ ensure that all $\SU(3)_F$ and $\U(1)_{B-L}$ anomalies cancel, respectively.\footnote{Assigning charges $\pm 2$ to $\mb{S}_{B-L}^\pm$ allows $\mb{N_{B-L}^c}$ to get a Majorana mass when $\mb{S}_{B-L}^-$ gets a vev. However, a complete model of flavor needs additional field content beyond those in  \Tab{tab:reps}, including a $\mb{6}$ to give a Majorana mass to $\mb{N_F^c}$ and a $\mb{\overline{6}}$ to generate the lepton Yukawas. See \Ref{Craig:2012di} for further discussion.}

There are a number of different options for how to generate the SM Yukawa couplings.  For pedagogical purposes, we will choose a structure that allows us to clearly delineate the role played by the different gauge groups in $G_{\rm aux}$ in generating the soft mass spectrum.  In the quark sector, we assume that the following dimension six operators arise after integrating out heavy vector-like fields:
\be
\label{eq:Wchoice}
\mb{W} \supset \frac{1}{\Lambda^2_{u}} \mb{S^-_H S_u H_u Q U^c} + \frac{1}{\Lambda^2_{d}} \mb{S^+_H S_d H_d Q D^c} .
\ee
Here, the up-type Yukawa matrix comes from $\vev{S^-_H S_u}/\Lambda^2_{u}$ and the down-type Yukawa matrix comes from $\vev{S^+_H S_d}/\Lambda^2_{d}$.  Instead of \Eq{eq:Wchoice}, we could have considered a more economical model where the $\mb{S_u}$ and $\mb{S_d}$ fields are charged under both $\SU(3)_F$ and $\U(1)_H$, allowing the Yukawa couplings to arise from dimension five operators.\footnote{\label{footnote:bothcharged}In this case, the $S_{u,d}$ vevs lead to mixing between the $\SU(3)_F$ and $\U(1)_H$ generators, giving the breaking pattern $\SU(3)_F \times \U(1)_H \rightarrow \SU(2)' \times \U(1)' \rightarrow \U(1)'' \rightarrow 0$.  The resulting soft mass spectrum contains mixed contributions proportional to $\alpha_H \alpha_F$, which is interesting but inconvenient for pedagogical purposes.}  Note that $\mb{S_{B-L}^\pm}$ need not play a role in generating the Yukawa couplings, though, due to the charges chosen, it can be used to generate right-handed neutrino masses.  If we only gauge a subset of $G_{\rm aux}$, then we can set the corresponding field in \Eq{eq:Wchoice} to a constant value.\footnote{For example, if $\U(1)_H$ is gauged but $\SU(3)_F$ is not, then we can use the simpler superpotential
\be
\label{eq:altWchoice}
\mb{W} = \frac{\lambda_u}{\Lambda_{u}} \mb{S^-_H H_u Q U^c} + \frac{\lambda_d}{\Lambda_d} \mb{S^+_H H_d Q D^c} ,
\ee
where $\lambda_u$ and $\lambda_d$ are proportional to the SM Yukawa matrices, avoiding the need to dynamically generate the hierarchical $S_{u,d}$ vevs.}

Given the superpotential in \Eq{eq:Wchoice}, the pattern of $\SU(3)_F$ gauge boson masses is determined by the measured flavor parameters.  We will make the simplifying assumption that $\langle \mb{S_u} \rangle \gg \langle \mb{S_d} \rangle$, such that the flavor boson mass-spectrum is dominated by the up-quark Yukawa. After performing a global $\SU(3)_F$ rotation we can diagonalize the flavor breaking matrices and denote
\be
\langle \mb{S_u} \rangle = \left(\begin{array}{ccc}v_{u1} & 0 & 0 \\0 & v_{u2} & 0 \\0 & 0 & v_{u3}\end{array}\right), \qquad \langle \mb{S_d} \rangle = V_{\rm{CKM}} \left(\begin{array}{ccc}v_{d1} & 0 & 0 \\0 & v_{d2} & 0 \\0 & 0 & v_{u3} \end{array}\right) V^T_{\rm{CKM}}.
\ee
This leads to the hierarchical flavor breaking pattern $\SU(3)_F  \rightarrow \SU(2)_F \rightarrow \emptyset$ where the flavor boson masses are
\begin{align}
M_V^2 [\sim \SU(3)_F/\SU(2)_F] & =  4 \pi \alpha_F  \left\{ \tfrac{8}{3} v_{u3}^2, (v_{u3}+v_{u2})^2,v_{u3}^2,v_{u3}^2,(v_{u3}-v_{u2})^2 \right\} , \\
M_V^2 [\sim \SU(2)_F] & =   4 \pi \alpha_F  \left\{ 2 v_{u2}^2, v_{u2}^2, v_{u2}^2 \right\} .
\end{align}
Explicitly inputting both the up-quark and down-quark Yukawa couplings, taking $\Lambda_{u} = \Lambda_d$ for simplicity ($\alpha = 1$ in the language of \Ref{Craig:2012di}), and denoting $v_{u3} \equiv v_F$, we have the flavor boson mass spectrum
\begin{align}
M_V^2 [\sim \SU(3)_F/\SU(2)_F] & \approx  4 \pi \alpha_F v_F^2  \left\{ 2.67, 1.02, 1.00, 1.00, 0.99 \right\}, \\
M_V^2 [\sim \SU(2)_F] & \approx   4 \pi \alpha_F v_F^2   \left\{  11.0, 5.60 , 5.55 \right\} \times 10^{-5},
\end{align}
clearly demonstrating the hierarchical symmetry breaking pattern for $\SU(3)_F$.

For the $\U(1)_{B-L}$ and $\U(1)_H$ gauge bosons, their masses are determined by the vevs  $\langle \mb{S_{B-L}^\pm} \rangle = v^\pm_{B-L}$ and $\langle \mb{S_H^\pm} \rangle = v^\pm_{H}$:
\begin{align}
M_V^2 [\U(1)_{B-L}] & =  32 \pi \alpha_{B-L} \left(({v^+_{B-L}})^2+({v^-_{B-L}})^2\right) , \\
M_V^2 [\U(1)_H] & =  8 \pi \alpha_H \left(({v^+_{H}})^2+({v^-_{H}})^2\right) .
\end{align}
With the chosen field content, we can freely adjust the masses of the $\SU(3)_F$, $\U(1)_{B-L}$, and $\U(1)_H$ gauge bosons.

\subsection{Soft Terms in Auxiliary Gauge Mediation}
\label{sec:auxsoft}

Once we choose $G_{\rm aux}$ representations for the messenger fields $\mb{\Phi}$, the soft terms in auxiliary gauge mediation follow directly from the general formulas in \Sec{sec:higgsedmediation}.   The Dynkin index of $\mb{\Phi}$ under $\SU(3)_F$ is $C(\mb{\Phi})$, and $\mb{\Phi}$ has charge $p_\Phi$ ($q_\Phi$) under $\U(1)_{B-L}$ ($\U(1)_H$).
We denote
\be
\delta_i  \equiv \left( \frac{{M_{V_i}}}{M} \right)^2,
\ee
where $M_{V_i}$ is the mass of the appropriate gauge superfield ($\SU(3)_F$, $\U(1)_H$, or $\U(1)_{B-L}$), and the generators $T^a$ always correspond to the $\SU(3)_F$ generators in the gauge boson mass eigenstate basis.  The soft terms are then given at the effective messenger scale (see \Sec{sec:scalarmasses}), and must be RG evolved down to the weak scale.

Using the results of \Sec{sec:scalarmasses}, the Higgs soft masses are given by
\be
\widetilde{m}_{H_u,H_d}^2 = q_\Phi^2 \frac{\alpha_H^2}{(2 \pi)^2}  \left | \frac{F}{M} \right |^2  f(\delta_H).
\label{eq:HiggsSoft}
\ee
The squark and slepton soft masses are given by
\be
\left(\widetilde{m}_q^2 \right)_{ij} = C(\mb{\Phi}) \frac{\alpha_F^2}{(2 \pi)^2}  \left | \frac{F}{M} \right |^2  \sum_a f(\delta_F^a) \, (T_q^a T_q^a)_{\{ij\}} + \eta \, p_\Phi^2 \frac{\alpha_{B-L}^2}{(2 \pi)^2}  \left | \frac{F}{M} \right |^2  f(\delta_{B-L}) \delta_{ij},
\label{eq:nonab}
\ee
where $\eta = 1$ for sleptons and $1/9$ for squarks, and $\{ij\}$ indicates that these indices have been symmetrized. As noted in \Ref{Craig:2012di}, the assumption that the up-quark Yukawa dominates implies that the off-diagonal terms in the squark and slepton mass matrices in the gauge interaction basis are extremely small, so as to be irrelevant for flavor constraints.

Next, applying the results from \Sec{sec:bmu} for the MSSM $B_\mu$ term:
\be
B_\mu = 2\mu_H q_\Phi^2 \frac{\alpha_H^2}{(2 \pi) ^2}  \frac{F}{M}  h(\delta_H),
\label{eq:bmu}
\ee
where the $\mu_H$ is the Higgsino mass.  We can similarly calculate the $A$-terms.  The holomorphic $h_u \widetilde{t}_L \widetilde{t}_R$ coupling is
\begin{eqnarray}
A_{h_u \widetilde{t}_L \widetilde{t}_R} & =&  \frac{ \lambda_t}{(2 \pi)^2} \Bigg(2 C(\mb{\Phi}) \alpha_F^2  \sum_a h(\delta^a_F) \, (T_q^a T_q^a)_{33} \nonumber \\ 
& & + \frac{2}{9} p_\Phi^2 \alpha_{B-L}^2 h(\delta_{B-L})  + q_\Phi^2 \alpha_H^2   h(\delta_H)  \Bigg) \left ( \frac{F}{M} \right ).
\label{eq:aterm}
\end{eqnarray}
Even though the messengers are charged under all factors of $G_{\rm aux}$, there are no crossterms containing e.g.\ $\alpha_H \alpha_F$.  This can be seen directly from the field rescalings, \Eqs{eq:rescaleab}{eq:rescalenonab}, which give rise to the $A$-terms.

Finally, we have the gaugino masses at three loops from \Sec{sec:gauginomasses}. Summing over all visible-sector fields in \Eqs{eq:gauginomassab}{eq:gauginomassnonab}, we have the gluino, wino, and bino masses
\begin{align}
\widetilde{M}_{\widetilde{g}} &= \frac{\alpha_S}{4 \pi^3}  \frac{F}{M} \left ( \frac{1}{2} C(\mb{\Phi}) \alpha_F^2  \sum_{a} h(\delta^a_F) + \frac{1}{3} p_\Phi^2  \alpha_{B-L}^2 h(\delta_{B-L}) \right ),\\
\widetilde{M}_{\widetilde{W}} &= \frac{\alpha_W}{4 \pi^3}  \frac{F}{M}  \left( \frac{1}{2} C(\mb{\Phi}) \alpha_F^2 \sum_{a} h(\delta^a_F)  + \frac{1}{2} q_\Phi^2 \alpha_H^2   h(\delta_H)  + 4 p_\Phi^2  \alpha_{B-L}^2 h(\delta_{B-L}) \right),\\
\widetilde{M}_{\widetilde{B}} &= \frac{\alpha_Y}{4 \pi^3}  \frac{F}{M}  \left(\frac{5}{6} C(\mb{\Phi}) \alpha_F^2 \sum_{a} h(\delta^a_F) + \frac{1}{2} q_\Phi^2 \alpha_H^2   h(\delta_H) + \frac{23}{9} p_\Phi^2  \alpha_{B-L}^2 h(\delta_{B-L}) \right) , \label{eq:bino}
\end{align}
where the prefactors from the $\SU(3)_F$ contribution come from the fact that all quark superfields are flavor fundamentals and have Dynkin index $1/2$.  Note that the gluino mass does not depend on $\alpha_H$ at this order, and we may exploit this freedom to obtain non-standard gaugino spectra.\footnote{Due to matter charged under both gauge groups, hypercharge may mix kinetically with $\U(1)_H$ and/or $\U(1)_{B-L}$, and gaugino mass-mixing may also occur.  However, one can show that even if this mixing is present the bino mass is still given by \Eq{eq:bino}.}

The various soft terms at the messenger scale in auxiliary gauge mediation, in particular the gaugino masses, are considerably different from those in standard gauge mediation. In auxiliary gauge mediation, the gaugino masses $\widetilde{M}$ are suppressed by two loops compared to the scalar masses $\widetilde{m}$, as opposed to standard gauge mediation where gauginos obtain mass at one loop and $\widetilde{M} \sim \widetilde{m}$.  For $\alpha_H = \alpha_{B-L} = 0$ we have the familiar GUT-motivated gaugino masses hierarchy at the messenger scale, $\MG : \MW : \MB = \alpha_S : \alpha_W : \alpha_1$, where $\alpha_1 = \frac{5}{3}\alpha_Y$ is the GUT-normalized hypercharge coupling. However, by turning on $\alpha_H$ and $\alpha_{B-L}$, we can change the hierarchy among the gaugino masses at the messenger scale and the wino or bino may end up closer in mass to (or even heavier than) the gluino.

\subsection{Renormalization Group Evolution}
\label{sec:RGevolution}
The above soft terms are the values at the effective messenger scale $\text{min} \{M_V,M\}$, which then must be RG evolved to the weak scale to determine the resulting phenomenology.  The RG behavior of the soft terms has important implications for the mini-split spectrum, particularly for the Higgs and third-generation squarks, which we will focus on here.  In the benchmark studies below, we perform the RG evolution of all soft parameters numerically.

In the MSSM, the RG equations for the third-generation squark masses and up-type Higgs masses contain the following terms \cite{ArkaniHamed:1997ab,Agashe:1998zz}:
\begin{itemize}
\item a one-loop term proportional to squared gaugino masses $\M_A^2$;
\item a two-loop term proportional to the first- and second- generation scalar masses-squared;
\item a one-loop hypercharge $D$-term $\alpha_Y Y_i \Tr(Y \m^2)$; and
\item a one-loop term proportional to
\be
X_t = |\lambda_t|^2(\m_{H_u}^2 + \m_{\Stop_R}^2 + \m_{\Stop_L}^2) + |\At|^2.
\ee
\end{itemize}
In auxiliary gauge mediation, the gaugino squared masses $\M_A^2$ appear formally at six loops and are therefore negligible in the RG evolution.   As has been pointed out previously in \Refs{Ibarra:2005vb,Arvanitaki:2012ps}, this absence of the gaugino contribution to the sfermion beta functions can allow the stops to run tachyonic at the weak scale. The two-loop term only contributes above the scale $\overline{\mu} \approx m_{1,2}$, but if $m_{1,2} \gg m_{3,i}$, this term can also push the stops tachyonic \cite{ArkaniHamed:1997ab,Agashe:1998zz}.   

Therefore, it is non-trivial to have a mini-split spectrum with the desired vacuum structure after RG evolution of the soft parameters.  In the case of auxiliary gauge mediation, the leading RG equation for the third-generation scalar soft masses and up-type Higgs in auxiliary gauge mediation is
\be
\frac{d \m_{H_{u}}^2}{d \log \overline{\mu}} = \frac{3}{8\pi^2} X_t, \qquad  \frac{d \m_{\Stop_{R}}^2}{d \log \overline{\mu}} = \frac{2}{8\pi^2} X_t, \qquad \frac{d \m_{\Stop_{L}}^2}{d \log \overline{\mu}} = \frac{1}{8\pi^2} X_t.
\label{eq:scalarRGE}
\ee
Compared to the full RG equation, we have kept only the $X_t$ term since the the hypercharge $D$-term vanishes at the messenger scale, and as long as $m_{1,2} \simeq m_{\Stop}$, the two-loop term can also be neglected.\footnote{The two-loop term only contributes when the first and second generation are moderately split from the third.
} Ignoring also the running of $\lambda_t$ and $\At$, we can find an analytic solution to the RG equation in \Eq{eq:scalarRGE}: 
\begin{align}
\m_{H_u}^2(\overline{\mu}) & = \m_{H_u}^2(M) - \frac{3\lambda_t^2}{8\pi^2}\left (\frac{|\At|^2}{\lambda_t^2} + \m_{H_u}^2(M) + 2\m_{\Stop}^2(M) \right )\log \frac{M}{\overline{\mu}},  \label{eq:RGonea}\\
\m_{\Stop_R}^2(\overline{\mu}) & =  \m_{\Stop}^2(M) -\frac{2\lambda_t^2}{8\pi^2} \left (\frac{|\At|^2}{\lambda_t^2} +  \m_{H_u}^2(M)  + 2 \m_{\Stop}^2(M) \right)\log \frac{M}{\overline{\mu}}, \label{eq:RGoneb}\\
\m_{\Stop_L}^2(\overline{\mu}) & = \m_{\Stop}^2(M) - \frac{\lambda_t^2}{8\pi^2}\left ( \frac{|\At|^2}{\lambda_t^2} + \m_{H_u}^2(M)  + 2 \m_{\Stop}^2(M) \right) \log \frac{M}{\overline{\mu}}. \label{eq:RGonec}
\end{align}
Here $\m_{\Stop}^2(M) \equiv \m_{\Stop_L}^2(M) = \m_{\Stop_R}^2(M)$ since both stops have the same soft mass at the messenger scale. We see that by adjusting $\m_{H_u}^2$ to be small enough compared to $\m_{\Stop}^2$ at the messenger scale, we can always arrange for $\m_{H_u}^2$ to run tachyonic and trigger electroweak symmetry breaking while $\m_{\Stop_{L,R}}^2$ remains positive.  Since the stop soft masses are controlled by the $\SU(3)_F \times \U(1)_{B-L}$ groups while the Higgs masses is controlled by $\U(1)_H$, there is ample parameter space where this occurs.\footnote{If we had $S_u$ and $S_d$ fields charged both under $\SU(3)_F$ and $\U(1)_H$ as in footnote~\ref{footnote:bothcharged}, then there would be mixed contributions proportional to $\alpha_F \alpha_H$.  In that case, one may have to rely more on the $\U(1)_{B-L}$ contribution to the stop masses to find viable parameter space.}

\section{Benchmark Scenarios}
\label{sec:examples}

\begin{table}[t]
\centering
\begin{tabular}{ c || c | c | c | c | c} \hline
\hline Benchmark & Low Scale & High Scale & Flavored & $B-L$ & superWIMP\\
\hline \hline $M_{\rm eff}$~[GeV] & $10^{10}$ & $10^{15}$ & $10^{10}$ & $10^{10}$ & $6\times10^{12}$\\
 $F/M$~[GeV] & $2\times10^5$ & $4\times10^5$ & $1\times10^5$ & $4\times10^5$ & $1\times10^6$\\
 \hline
 $\sqrt{C(\mb{\Phi})}\, \alpha_F$ & 0.9 & 0.9 & 2.5 & --- & 0.6\\
 $\delta_F$ & 0.1 & 0.1 & 260 & --- & 0.1\\
\hline
 $p_\Phi \, \alpha_{B-L}$ & --- & --- & --- & 3.0 & 0.8\\
 $\delta_{B-L}$ & --- & --- & --- & 0.1 & 0.1\\
\hline
 $q_\Phi \, \alpha_H$ & 0.9 & 0.9 & 0.4 & 0.6 & 0.6\\
  $\delta_H$ & 0.1 & 0.1 & 0.1 & 0.02 & 0.0125\\
  \hline
   $\tan\beta$ & 4.469 & 4.396 & 20.05 & 4.552 & 3.95\\
       $\mu_H$~[TeV] & 11.9 & 36.9 & 0.8 & 34.7 & 45.8\\
        $\sqrt{B_\mu}$~[TeV] & 18.3 & 45.6 & 1.5 & 35.4 & 67.3\\
    \hline
   $m_{3/2}$ [GeV] & $1.5 \times 10^{-3}$ & 300 &$7.6 \times 10^{-4}$ & $6.8 \times 10^{-3}$ & 1.9 \\

\hline\hline
\end{tabular}
\caption{Parameters for five auxiliary gauge mediation benchmark points: ``Low Scale'' with a low messenger mass, ``High Scale'' with a large messenger mass, ``Flavored'' with non-negligible splittings between the third-generation and first-two-generation scalars, ``$B-L$'' which employs only the $\U(1)_{B-L} \times \U(1)_H$ gauge groups, and a ``superWIMP'' scenario which can accommodate gravitino dark matter. In \textsc{SoftSUSY}, $\tan \beta$ is an input which sets the Higgsino mass $\mu_H$ after solving for electroweak breaking conditions.  The Higgs mass is 126 GeV for each benchmark, consistent with LHC results.  Except for $\tan \beta$, all of these values are specified at the effective messenger scale $M_{\rm eff} = \min\{M,M_V\}$ described in \Sec{sec:scalarmasses} and set the UV boundary condition for RG evolution to the weak scale.  For benchmarks where each factor of $G_{\textrm{aux}}$ has its own $\delta$, each soft term should really be run down from its corresponding effective messenger scale. However, since none of our benchmarks feature vastly different values of $\delta$, the error incurred by taking a single messenger scale for all soft terms (here taken to be the minimum of the various effective messenger scales) is small and does not significantly change the phenomenology.  }
\label{tab:BenchmarkParams}
\end{table}

\begin{figure}[t]
\begin{center}
\includegraphics[height=0.55\textwidth]{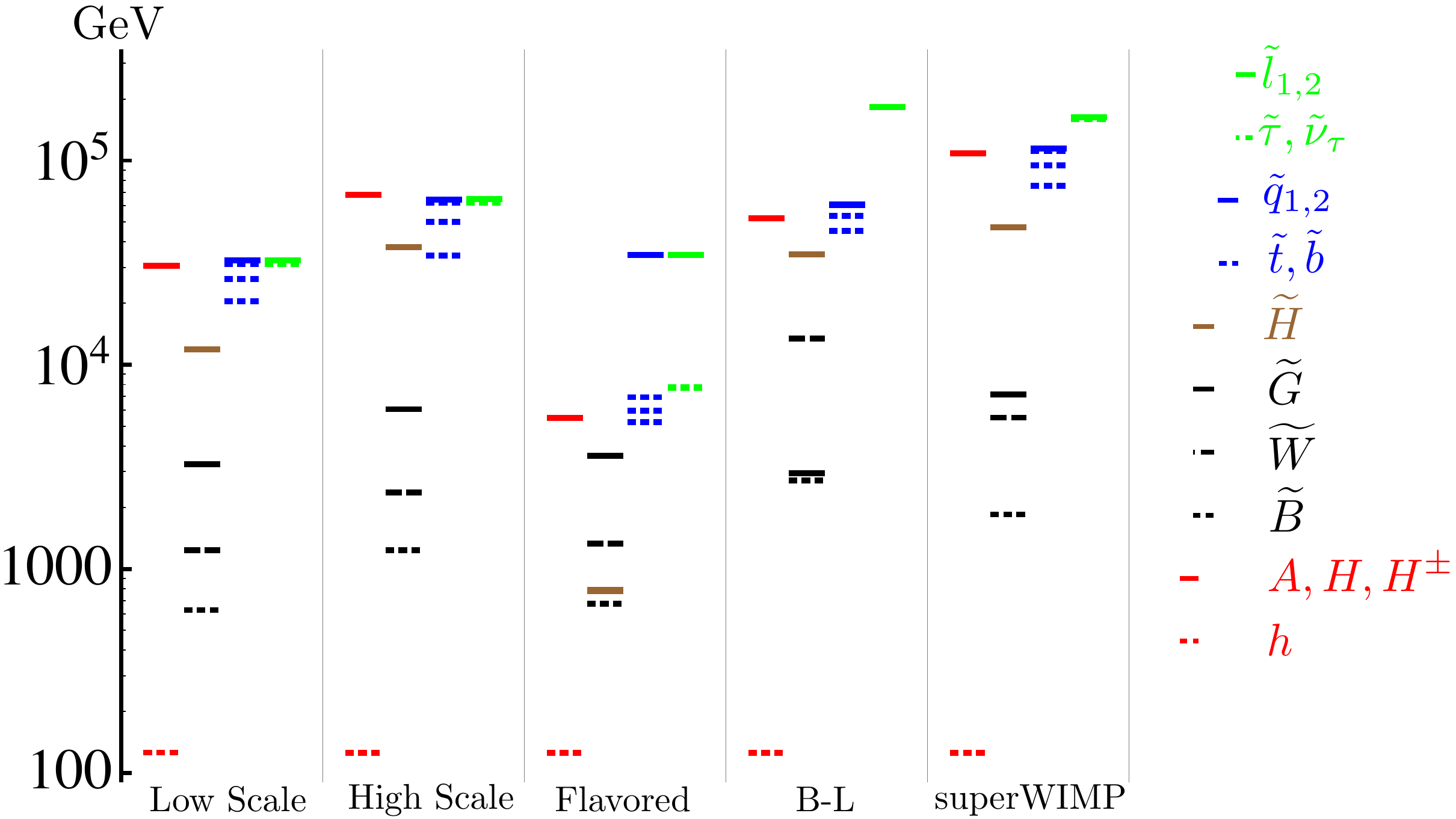}
\caption{Weak scale spectra for the five benchmark points specified in \Tab{tab:BenchmarkParams} and described in the text.  Each benchmark is split into four columns depicting (from left to right) Higgs sector scalars, inos, squarks, and sleptons.  In the third and fourth columns, third generation scalars are shown in dotted lines and first two generations in solid lines.}
\label{fig:BenchmarkSpectra}
\end{center}
\end{figure}

As proof of principle that auxiliary gauge mediation can generate a realistic mini-split spectrum, we present five benchmark points which result in a Higgs mass of approximately 126 GeV.  The messenger scale parameters for these benchmarks are given in \Tab{tab:BenchmarkParams}. The RG evolution to the weak scale is performed using $\textsc{SoftSUSY~3.3.8}$ \cite{Allanach:2001kg}, modified to allow the auxiliary gauge mediation boundary conditions at the messenger scale, and the resulting spectrum is shown in \Fig{fig:BenchmarkSpectra}.\footnote{It may well be the case that the operating accuracy of \textsc{SoftSUSY} is less than the fine-tuning required to achieve the electroweak symmetry breaking conditions and that additional uncertainty arises through the hierarchical RG thresholds.  However, we expect that the true physical spectrum is likely to be close enough to the spectrum given by \textsc{SoftSUSY} for the practical purpose of demonstrating the features of this setup.}  Phenomenological discussions of the benchmarks appear in the subsequent subsections.

In all of the benchmarks, the overall scale of the spectrum is set by requiring the gluino masses to be above 1.5 TeV, to ensure consistency with current collider bounds for scenarios where the lightest SUSY particle (LSP) is a gravitino \cite{CMS:2012kwa, Chatrchyan:2012bba,Aad:2012zza}.  For the auxiliary gauge couplings to remain perturbative, this requires $F/M \gtrsim 100$ TeV. This in turn places the sfermion mass scale at about $\widetilde{m}^2 \gtrsim (10^4~\GeV)^2$, which is precisely the required scale for a 126 GeV Higgs \cite{Arvanitaki:2012ps}. The Higgs soft masses are independent from the squark and slepton masses, since they depend only on $\alpha_H$ and not $\alpha_F$ or $\alpha_{B-L}$, but to ensure the vacuum does not break color we must have $\m^2_H \lesssim \m^2_3$ (see \Sec{sec:RGevolution} and \Sec{sec:minimal}).  The gravitino mass $m_{3/2}$ should be taken as a lower bound, since its mass could be lifted with multiple SUSY breaking \cite{Cheung:2010mc} or gravitino decoupling \cite{Luty:2002ff, Craig:2008vs}. 

As previously mentioned in the introduction, in any mini-split model there are two different types of tunings which one must be aware of.  The first tuning, which is widely appreciated, is the tuning of the Higgs sector parameters necessary to obtain a hierarchy between the electroweak symmetry breaking scale and the scalar soft masses.  In the case of auxiliary gauge mediation, the Higgsino mass $\mu_H$ is a free parameter which can be tuned for this purpose.   

The second tuning, not often discussed, is when one has to tune model parameters to precise values in order for the model to be viable.  This is the case, for example, if typical model parameters lead to color-breaking vacua or if the model generically leads to inappropriate values for $B_\mu$. Our models avoid this second type of tuning, with only the first type of tuning which is irreducible in mini-split models. Indeed, in the benchmarks discussed here, only one parameter needs to take finely adjusted values, and the mini-split spectrum, including an acceptable Higgs sector, can be accommodated within much of the parameter space of the model.  
 
\subsection{Two $\SU(3)_F \times \U(1)_H$ Models}

Our first two benchmarks utilize just the $\SU(3)_F \times \U(1)_H$ subgroup of $G_{\rm aux}$ to mediate SUSY breaking.  Here, squarks and sleptons of a given generation receive identical soft masses from the $\SU(3)_F$ mediation.  The gluino obtains mass at three loops from diagrams involving just the $\SU(3)_F$ gauge group, whereas the wino and bino feel two loop contributions from both gauge groups.  Thus the ratio of gaugino masses is different from those found in other scenarios such as anomaly or gauge mediation.  In particular, it is possible for the mass of the bino and wino to be raised closer to the gluino than in other models.

We consider two benchmark scenarios:   ``Low Scale'' with a relatively low messenger masses, and ``High Scale'' with a higher messenger mass scale.  We take $\delta_F \lesssim 1$ such that the generation-dependent splitting is small, and all the squark and slepton generations obtain similar soft masses at the messenger scale.  These scenarios economically realize the ``mini-split'' spectrum.  
There is some small splitting of generations, particularly due to the running of the stop mass, however the scalars all have mass beyond the LHC reach of $\widetilde{m} \gtrsim 10$ TeV.  The Higgsinos are also reasonably heavy, requiring smaller values of $\tan \beta \sim 5$.  Both of these scenarios would lead to generic mini-split LHC phenomenology, with gluinos decaying through off-shell squarks in a decay chain which terminates with an invisible gravitino.  Displaced vertices could potentially arise from bino decays.

A feature of this scenario compared to other mini-split models is that by including the $\U(1)_H$ symmetry, the appropriate Higgs sector soft parameters, including $B_\mu$, can be generated without requiring additional couplings between the Higgs and SUSY-breaking sectors.

\subsection{A Flavored $\SU(3)_F \times \U(1)_H$ Model}

Taking the same $\SU(3)_F \times \U(1)_H$ subgroup, we can realize a ``Flavored'' benchmark point by taking $\delta_F \gtrsim 1$. In this case, flavor mediation generates greater masses for the first and second generation scalars, with third generation scalar masses somewhat suppressed, as described in \Ref{Craig:2012di}.  This can make for novel mini-split spectra with some smoking gun phenomenological features.  For the ``Flavored'' benchmark point we choose a large value of $\delta_F$ such that the third-generation squark mass is suppressed by a factor $\sim 6$ relative to the first-two-generation squarks.  Since the gluino decays proceed via off-shell squarks this would lead to extremely top- and bottom-rich gluino decays, with third-generation decays a factor $6^4$ more frequent than decays involving the first-two-generation squarks.  Top- and bottom-tagging would then enhance the LHC sensitivity to such flavored mini-split scenarios.  Another notable feature of this scenario is that, since the $\SU(3)_F$ gauge symmetry treats sleptons and squarks equally (a feature demanded by anomaly-cancellation) any flavored spectrum automatically keeps the sbottoms and staus light, alongside the stop.

This flavored benchmark also features reasonably light higgsinos, with $m_{\widetilde{H}} \sim 750$ GeV and a larger value of $\tan \beta \sim 20$.  Such light Higgsinos are possible as $m_{H_u}^2$ can be tuned small if the amount of running is tuned.  Then to obtain electroweak symmetry breaking a smaller $|\mu_H|^2$ can be tuned against $m_{H_u}^2$, leading to Higgsinos significantly lighter than the squarks and sleptons, although this is not specific to the auxiliary gauge mediation scenario.

\subsection{A $\U(1)_{B-L} \times \U(1)_H$ Model}

Another interesting scenario to consider is whenever the mediation is entirely flavorless, such that gauge mediation only occurs via the $\U(1)_{B-L} \times \U(1)_H$ subgroup.  Mediation via a $\U(1)_{B-L}$ symmetry was previously considered in \Ref{Arvanitaki:2012ps} for generating a mini-split spectrum.  However, in order to generate Higgs soft parameters this gauge symmetry had to be significantly mixed with $\U(1)_Y$, with the mixing parameter taking a specific value to avoid color-breaking vacua.  These issues are circumvented here simply by employ the $\U(1)_H$ symmetry, which can generate Higgs sector soft masses and the $B_\mu$ term at the appropriate scale.

The ``$B-L$'' benchmark has some very interesting features, which can be traced back to the fact that squarks carry $\U(1)_{B-L}$ charge which is three times smaller than sleptons.  The first obvious feature is that sleptons tend to have masses a factor $\sim 3$ larger than squarks.  This would also further suppress leptonic high intensity probes.  This is in sharp contrast to the situation in standard gauge mediation, where the squarks are several times heavier than the sleptons, as well as in the hypercharge-mixed mini-split model of \Ref{Arvanitaki:2012ps}.

A less immediate consequence follows from the fact that gluino soft masses are mediated via loops involving squarks, whereas the winos and bino also obtain contributions from loops of sleptons.  Due to the larger slepton $\U(1)_{B-L}$ charge, the bino and wino masses can be raised significantly, close to, or above the gluino mass.  This is demonstrated in \Fig{fig:BenchmarkSpectra} where the wino is much heavier than the gluino, and the bino and gluino are almost degenerate.  Such gaugino mass patterns are rather unique and do not arise in ordinary gauge-mediated realizations of mini-split. In \Sec{sec:minimal}, we show how the same gross features can arise in a more economical model with a single mediating $\U(1)$ gauge group.

\subsection{SuperWIMPs from $\SU(3)_F \times \U(1)_{B-L} \times \U(1)_H$}

Our final benchmark employs all three factors of $G_{\rm aux}$, and was chosen to realize the superWIMP scenario \cite{Feng:2003xh, Feng:2003uy} discussed in \Ref{Feng:2012rn}.  The ``SuperWIMP'' benchmark has a gravitino mass of $1.9~\GeV$ and a bino mass of $1.6~\TeV$.  In gauge mediation with only a single SUSY-breaking sector, the gravitino is almost always the LSP, but once the the gravitino is heavy enough to be a viable cold dark matter candidate, gravity-mediated contributions to SUSY breaking can pollute the flavor-blind gauge-mediated soft terms and cause flavor problems. One solution is to have the current relic abundance of gravitino dark matter be produced non-thermally, through the decay of a long-lived WIMP after freeze-out. In gauge mediation, the bino typically plays the role of the WIMP and a light gravitino can be a superWIMP. Indeed a gravitino LSP and bino NLSP of the appropriate masses can also satisfy conditions on the bino lifetime from big bang nucleosynthesis and ensure that small-scale structure formation is not disrupted by free-streaming gravitinos. A full analysis of these cosmological constraints is beyond the scope of this paper, but we note that the preferred parameter space (gravitino at $1-10~\GeV$, bino at $1-5~\TeV$) given in \Ref{Feng:2012rn} is easily accommodated in our model. 

\section{A Minimal Mini-Split Model}
\label{sec:minimal}
The examples of \Sec{sec:examples} demonstrate a wide variety of possibilities for mini-split model building with auxiliary gauge mediation.  Motivated by minimality, it is interesting to consider the smallest gauge symmetry required to generate a mini-split spectrum with the correct SM vacuum.  In this case the auxiliary gauge group is some subgroup of the full available symmetry which, requiring appropriate Higgs sector soft terms and masses for colored superpartners, is 
\be
\U(1)_{X \equiv B-L+k H} \subset \U(1)_{B-L} \times \U(1)_H.
\ee
Here $k$ denotes the freedom to choose the normalization of the Higgs charges relative to $B-L$ charges.  The parameter $k$ is not entirely free as there are constraints on the charge of Higgs fields from RG evolution. From Eqs.~(\ref{eq:RGonea})--(\ref{eq:RGonec}) it is clear that to have radiative EW symmetry breaking and a color-preserving vacuum one requires $2 \m_{H_u}^2 \lesssim 3 \m_{\Stop}^2$ at the messenger scale (assuming small $A$-terms and only considering one-loop running).  For the $\U(1)_X$ symmetry considered above, choosing the overall normalization by setting the usual baryon charge $q_q = 1/3$ constrains $q_H^2 \lesssim 1/6$.  As long as this criterion is satisfied, there is no barrier to constructing a minimal model of auxiliary gauge mediation based on this single $\U(1)_X$ gauge symmetry, with the understanding that the MSSM Yukawa couplings are generated as in \Eq{eq:altWchoice} and a separate spurion may be responsible for the generation of Majorana neutrino masses.

\begin{figure}

\begin{floatrow}
\capbtabbox{%
   \begin{tabular}
   { c || c} \hline
\hline Benchmark & Minimal Model\\
\hline \hline $M_{\rm eff}$~[GeV] & $10^{10}$ \\
 $F/M$~[GeV] & $7\times10^5$ \\
 \hline
 $q_\Phi \, \alpha_X$ & 3.0 \\
  $\delta_X$ & 0.04\\
  \hline
   $\tan\beta$ & 3.045\\
       $\mu_H$~[TeV] & 51.5 \\
       $\sqrt{B_\mu}$~[TeV] & 88.3 \\
    \hline
   $m_{3/2}$ [GeV] & $5.3 \times 10^{-3}$ \\

\hline\hline
\end{tabular}
}{
\caption{Parameters for the minimal auxiliary gauge mediation model with a single $\U(1)_X$ gauge symmetry with lepton, quark, and Higgs charges $q_l = 1$ and $q_q = q_H = 1/3$.}
  \label{tab:min}%
}
\ffigbox{%
  \includegraphics[height=0.4\textwidth]{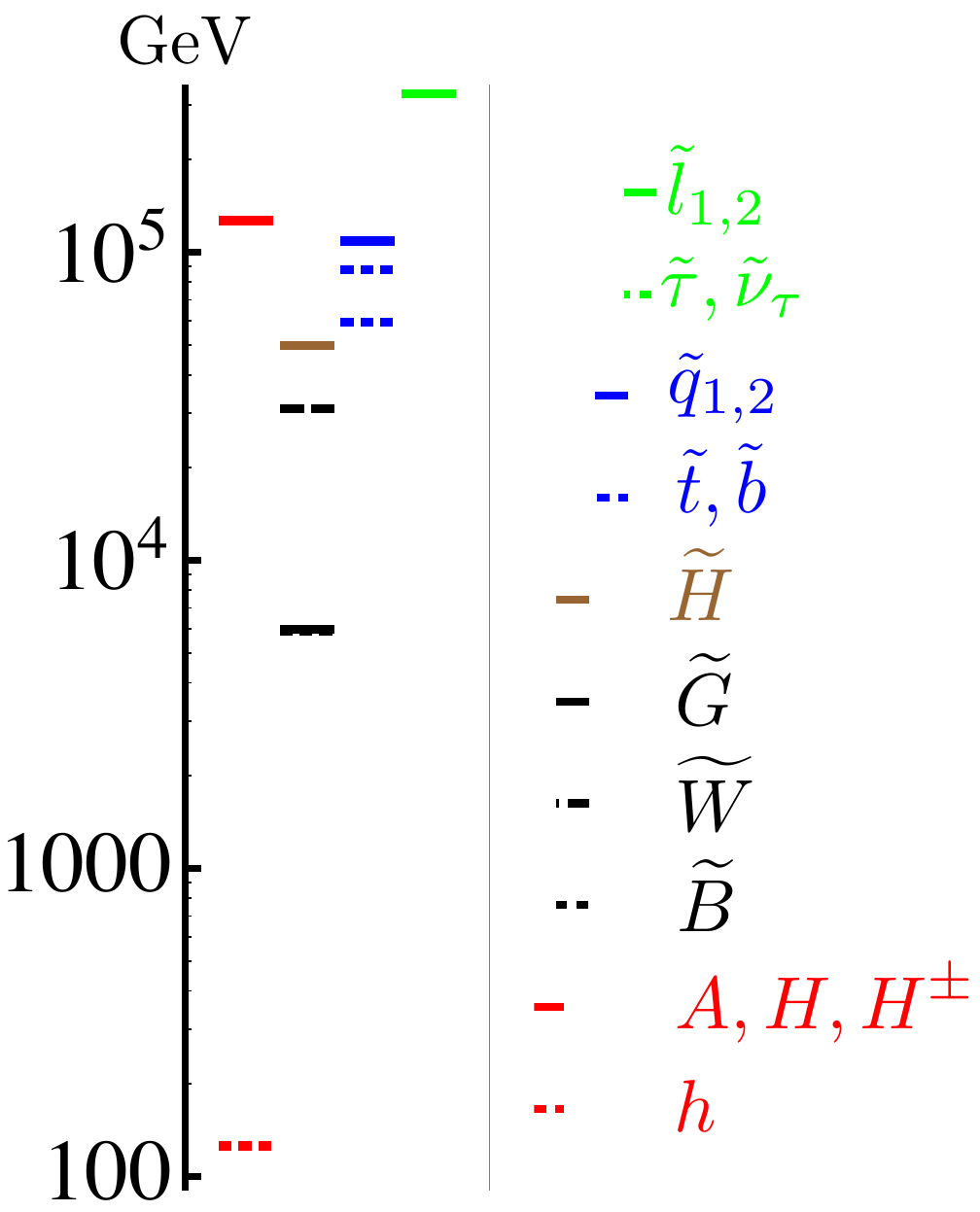}%
}{
  \caption{Particle spectra for the minimal $\U(1)_X$ auxiliary gauge mediation model.  Conventions follow \Fig{fig:BenchmarkSpectra}.  Due to the $B-L$ nature of the squark and slepton charges the sleptons are a factor $\sim3$ more massive than squarks.  The wino is the heaviest of the gauginos due to the large three-loop contributions involving sleptons.   The gluino and bino happen to be close in mass for this benchmark.}%
  \label{fig:min}
}

\end{floatrow}
\end{figure}

As an example minimal scenario, consider $\U(1)_X$ where the lepton charge is $q_l = 1$ and the Higgs and quark charges are $q_H = q_q = 1/3$ (i.e.~$k = 1/3$).  We show a ``Minimal'' benchmark parameter choice in \Tab{tab:min} and the corresponding particle spectrum in \Fig{fig:min}.\footnote{Again, due to the inherent uncertainties introduced with such large fine-tuning, this spectrum should be taken as demonstrative of the overall qualitative features.}  As expected, the sleptons are heavier than the squarks by a factor $\sim3$, and due to large three-loop contributions from sleptons the wino and bino masses have increased relative to the gluino, leading to a non-standard gaugino spectrum.

A full study of this minimal auxiliary gauge mediation scenario is beyond the scope of this work.  However, this benchmark demonstrates that the full mini-split spectrum, with the necessary Higgs sector soft parameters and scalars two loop factors heavier than gauginos, can all be generated from a single $\U(1)$ gauge symmetry.

\section{Conclusions}
\label{sec:conclusions}

Naturalness has long been a guiding principle for constructing models of weak scale SUSY, but the observed Higgs boson at 126 GeV raises the possibility that some tuning of parameters might be necessary for successful electroweak symmetry breaking.  In this light, mini-split SUSY is an attractive scenario, and we have shown that a spectrum of heavy sfermions with light gauginos automatically arises in gauge mediation by the auxiliary group $G_{\rm aux} = \SU(3)_F \times \U(1)_{B-L} \times \U(1)_H$.  The key ingredient is the $\U(1)_H$ symmetry acting on the Higgs doublets, which generates the appropriate Higgs sector soft parameters (including $B_\mu$) such that only a single parameter needs to be tuned to have a viable spectrum.

The phenomenology of auxiliary gauge mediation shares many of the same features as generic mini-split models, with a few unique features.  The $\U(1)_H$ factor raises the masses of the bino and wino compared to standard scenarios, leading to lighter gluinos within phenomenological reach.  If $\SU(3)_F$ is present with $\delta^a \gtrsim 1$, then the third-generation sfermions are lighter than those of the first two generations, leading to gluino decays with top- and bottom-rich cascade decays.  Mediation with the $ \U(1)_{B-L}$ factor gives much larger masses to sleptons than squarks, and auxiliary gauge mediation with the full auxiliary group can give rise to superWIMP gravitino dark matter. Finally, we have shown that auxiliary gauge mediation with a single abelian group $\U(1)_{B-L+kH}$ can reproduce the gross features of a mini-split spectrum with the correct Higgs mass.

In our analysis, we have treated the breaking of $G_{\rm aux}$ and the mediation of SUSY breaking as independent modules, but it is attractive to consider the possibility that auxiliary gauge breaking and SUSY breaking might be more intimately related, since both can occur at intermediate scales.  Indeed, models with dynamical SUSY breaking often include spontaneously broken gauge symmetries \cite{Affleck:1984xz,Intriligator:2006dd}, some of which could be potentially be identified with $G_{\rm aux}$.  Given the model building challenge of generating the hierarchical $\SU(3)_F$ flavor breaking, it is encouraging that auxiliary gauge mediation with just $\U(1)_{B-L} \times \U(1)_H$ (or $\U(1)_{B-L+kH}$) is sufficient to generate a mini-split spectrum.  On the other hand, tying $\SU(3)_F$ breaking to SUSY breaking may give new insights into SM flavor.  More generally, auxiliary gauge mediation is a reminder that there can be rich dynamics in the ``desert'' between the weak scale and Planck scale, and these dynamics may leave their imprint in novel SUSY spectra.

\begin{acknowledgments}
We thank Nathaniel Craig for collaborating during the early stages of this work, Ben Allanach and Matthew Dolan for helpful conversations regarding SoftSUSY, and the anonymous referee for bringing a helpful reference to our attention. YK thanks Ian Low and the participants of TASI 2013 for stimulating discussion.  MM thanks Andrew Larkoski for conversations.  This work is supported by the U.S. Department of Energy (DOE) under cooperative research agreement DE-FG02-05ER-41360.  YK is supported by an NSF Graduate Fellowship, MM is supported by a Simons Postdoctoral Fellowship, and JT is supported by the DOE Early Career research program DE-FG02-11ER-41741.
\end{acknowledgments}

\appendix
\section{All-Orders Result for $A$-terms and $B$-terms}
\label{app:allorders}

\begin{figure}[t]
\begin{center}
\includegraphics[height=0.33\textwidth]{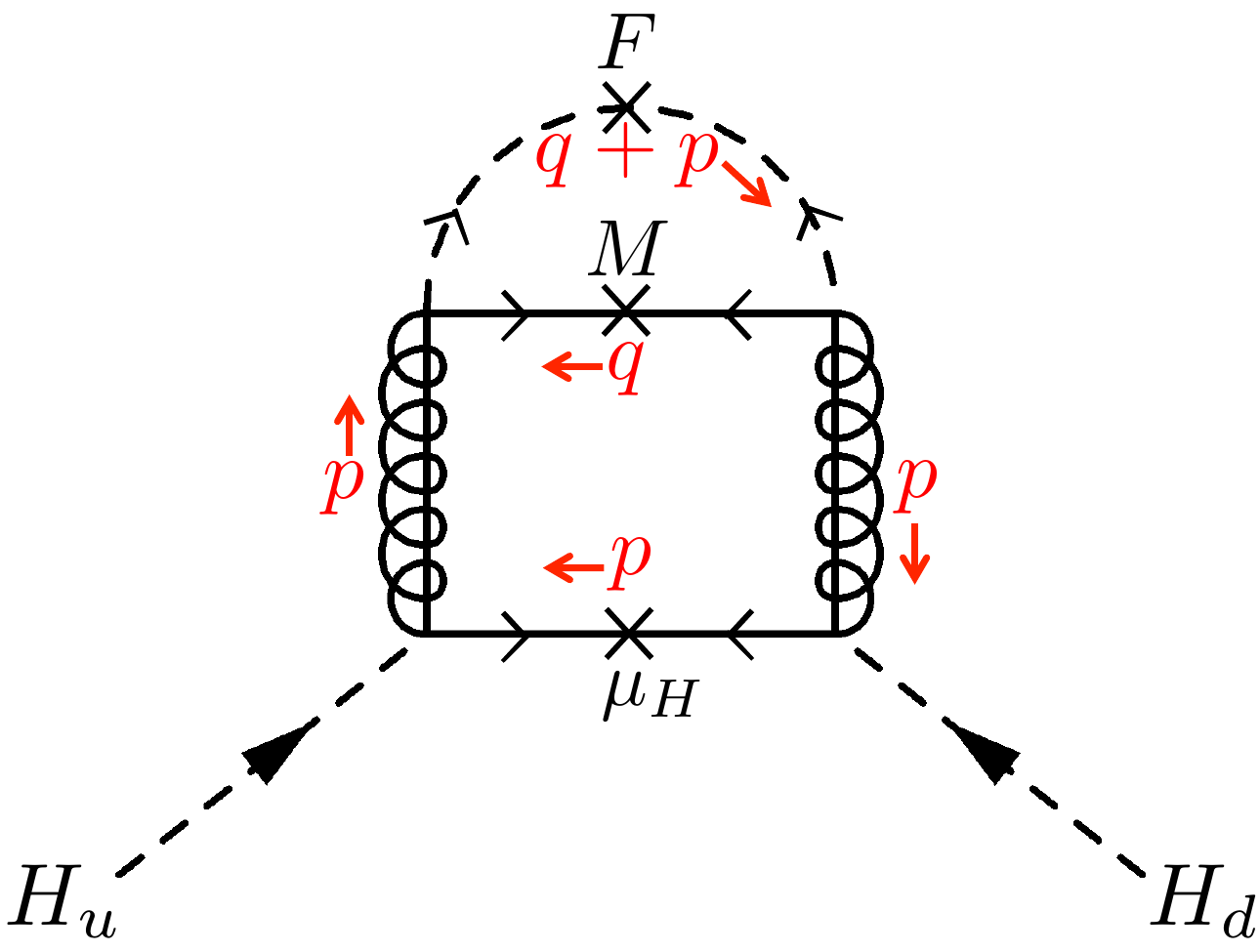}
\end{center}
\caption{Generation of $B_\mu$ at two loops from gauginos and messengers.  The diagram for $A$-terms is analogous, except with the Higgsino mass $\mu_H$ replaced by a scalar vertex.  The two-loop calculation performed here includes all orders in $F/M^2$, however the perturbative mass insertions for the messengers have been depicted here to demonstrate the chirality flips required for the generation of the lowest-order term. The red arrows show the momentum routing. }
\label{fig:twoloop}
\end{figure}

In this appendix, we present the first two-loop calculation of $A$- and $B$-terms to all orders in $F/M$ by a component Feynman diagram calculation. This calculation is simplified as only a single diagram contributes, shown in \Fig{fig:twoloop}.

\subsection{Result in Higgsed Gauge Mediation}
\label{app:allordersbroken}

We start with the case of a broken gauge group, where the diagram in \Fig{fig:twoloop} is finite. For the $B_\mu$ term, the result is
\be
B_\mu  = 16 \mu g_H^4 q_\Phi^2 MF  \, I (M_V, M, F),
\ee
where the familiar two loop integral is
\be
\label{eq:apploop}
I(M_V,M,F) = \int \frac{d^4 p \, d^4 q}{(2\pi)^8} \frac{1}{(p^2 - M_V^2)^2} \frac{1}{q^2 - M^2} \frac{1}{((q+p)^2 - (M^2 + F))((q+p)^2 - (M^2-F))}.
\ee
Here, $M_V$ is the gaugino mass, $M$ is the fermionic messenger mass, and $M^2 \pm F$ are the scalar messenger masses-squared.  After summing over the two scalar messenger mass eigenstates, the upper messenger loop gives the last factor in the loop integral of \Eq{eq:apploop}.  This finite integral can be evaluated by the usual method of Feynman parameters, giving
\be
B_\mu = 2 \mu_H  q_\Phi^2 \frac{\alpha_H^2}{(2\pi)^2} \frac{F}{M} \tilde{h}\left( \kappa,\delta \right),
\ee
where $\kappa = F/M^2$, $\delta = M_V/M$, and 
\be
\tilde{h}(\kappa,\delta)= \int_0^1 dw \int_0^1 dx \int_0^{1-x} dy \, \frac{2(1-w)}{w(1 + (x-y)\kappa) - (1-w)((x+y)^2 - (x+y))\delta}.
\ee
Making the change of variables $u = x+y$, $v = x-y$, two of the Feynman integrals can be evaluated analytically, giving
\begin{align}
\tilde{h}(\kappa,\delta ) = \frac{1}{\kappa}&\int_0^1 du \, \Bigg \{ \textrm{Li}_2\left(1+ \frac{1-\kappa u}{u(u-1)\delta} \right) - \textrm{Li}_2\left(1+ \frac{1+\kappa u}{u(u-1)\delta} \right) \\ \nonumber
& + \frac{\kappa \delta u^2(u-1)\log \left ( \frac{1-\kappa^2 u^2}{u^2(1-u)^2\delta^2} \right ) - 2(\delta(u-u^2) + \kappa^2 u^2 - 1)\tanh^{-1}(\kappa u)}{u^2 \kappa^2 - (1-(u-u^2)\delta)^2} \Bigg \}. 
\end{align}
For $\kappa = 0$, one can perform the $u$ integral analytically to show that $\tilde{h}(0,\delta)$ matches precisely with $h(\delta)$ given in \Eq{eq:hdelta}.  The $A$-terms lead to the same loop integrals and functional form for $\tilde{h}(\kappa,\delta )$.  

\subsection{Results in Standard Gauge Mediation}
\label{app:allordersunbroken}
To make contact with results from standard gauge mediation, the $A$- and $B$-terms must be determined for an unbroken mediating gauge group. In this case, the internal gauginos become massless in \Fig{fig:twoloop}, leading to an IR divergence which, although vanishing in physical observables, must be regulated to enable comparison with expressions for $A$-terms and $B$-terms in the literature.\footnote{It should be noted that the gauginos obtain mass at one-loop.  However, inserting this one-loop mass to regulate the two-loop diagram in \Fig{fig:twoloop} formally leads to a three-loop result, and is thus not included in the leading result, though they were included in the calculation of \Ref{Rattazzi:1996fb}. }  Formulae in the gauge mediation literature are often quoted using dimensional reduction with the minimal subtraction scheme, i.e.\ $\overline{\text{DR}}$.  Hence it makes sense to regulate the divergence in a way  which makes contact with the $\overline{\text{DR}}$ RG scale $\overline{\mu}$, allowing a comparison with the standard results for $A$- and $B$-terms in gauge mediation.

We regulate this IR divergence following the prescription used in e.g\ Eq.\ (2.21) of \Ref{Martin:2003qz}.\footnote{The specific integral regulated in this manner in \Ref{Martin:2003qz} is the same as each of the contributing integrals of \Fig{fig:twoloop} which are summed to give \Eq{eq:apploop}.  Hence the structure of the IR divergence is identical and we can employ the same prescription here.}  The regulated integral is evaluated as
\be
I(0,M,F) = 	\lim_{M_V\to0} \left[  I(M_V,M,F) + G(M,F) \log \left( \frac{M_V^2}{\overline{\mu}^2} \right) \right] ,
\ee
where $\overline{\mu}$ is the $\overline{\text{DR}}$ RG scale and $G(M,F)$ is the finite one-loop subintegral involving only messenger fields.  This cancels the logarithmic divergence in $M_V$ and, practically speaking, amounts to making the replacement $M_V\rightarrow \overline{\mu}$ in $I(M_V,M,F)$ and taking the limit $\overline{\mu} \to 0$. We obtain the final result
\be
B_\mu = 2 \mu_H  q_\Phi^2 \frac{\alpha_H^2}{(2\pi)^2} \frac{F}{M} h_{\overline{\text{DR}}},
\ee
where
\be
h_{\overline{\text{DR}}} = 1+\log \left( \frac{M^2}{\overline{\mu}^2} \right) ,
\ee
and similarly for $A$-terms as they arise from the same diagram.  Thus we find that in standard gauge mediation the $A$- and $B$-terms do not vanish at the messenger scale when the IR-divergent contributions are regulated with $\overline{\text{DR}}$.  Note that in \Ref{Rattazzi:1996fb}, the finite piece (which is regulator dependent) was absorbed into a redefinition of the messenger threshold, $M \to eM$.  However, if one uses $\overline{\text{DR}}$ then the messenger threshold really is $M$ and the finite piece is genuine. Furthermore we can make a direct connection with the analytic continuation methods developed in \Refs{Craig:2012yd,Craig:2012di} for an unbroken mediating gauge group.  This once again shows the consistency between the analytic continuation methods of \Refs{Craig:2012yd,Craig:2012di} and brute force Feynman diagram calculations, in this case for unbroken mediating gauge groups.

\bibliographystyle{JHEP}
\bibliography{MiniSplitBib}{}

\end{document}